\documentclass[a4paper,11pt]{article}

\usepackage{jhep-mod}
\usepackage[T1]{fontenc} 
\usepackage{blindtext}
\usepackage{tikz}
\usepackage[compat=1.0.0]{tikz-feynman}
\usepackage{amsmath}
\usepackage{soul}  

\usepackage[normalem]{ulem}
\usepackage{color}

\newcommand{\mO}{\mathcal{O}}

\newcommand*\dif{\mathop{}\!\mathrm{d}}

\tikzset{graviton/.style={decorate, decoration={snake, amplitude=.4mm, segment length=1.5mm, pre length=.5mm, post length=.5mm}, double}}

\title{\boldmath Gravitational Coleman-Weinberg Mechanism}

\author[a]{Clara~\'Alvarez-Luna~\footnote{Corresponding author.},}

\author[a]{Sergio~de~la~Calle-Leal,}

\author[a]{Jos\'e~A.~R.~Cembranos,}  

\author[a]{Juan~Jos\'e~Sanz-Cillero}

\affiliation[a]{Departamento de F\'isica Te\'orica and IPARCOS, Facultad de Ciencias F\'isicas,
\\Universidad Complutense de Madrid,
\\Plaza de las Ciencias 1, 28040 Madrid, Spain}

\emailAdd{c.a.luna@ucm.es}
\emailAdd{sergidel@ucm.es}
\emailAdd{cembra@fis.ucm.es}
\emailAdd{jjsanzcillero@ucm.es}

\abstract{
The Coleman-Weinberg mechanism provides a procedure by which a scalar field, which initially has no mass parameters, acquires a mass due to the anomalous nature of scale symmetry. Loop corrections trigger a spontaneous symmetry breaking and the appearance of a non-trivial vacuum. We first review the basic example of the Coleman-Weinberg mechanism, scalar Quantum Electrodynamics, in a perturbative regime where the scalar particle becomes massive through photon loops.
We then present the main results of this article, what we name the gravitational Coleman-Weinberg mechanism: we analyse the same effect in a gravitational theory without explicit energy scales at tree-level. Finally, we also study the mechanism for two scalar fields in the mentioned gravitational theory. We will derive the gravitational Coleman-Weinberg potentials, analyse the parameter space where we have a symmetry breaking, and obtain the value of the corresponding scalar masses.} 


\begin{document}
 
\maketitle
\flushbottom

\section{Introduction}

In 1973, theoretical physicists Sidney Coleman and Erick Weinberg published a work introducing what it has been known as the Coleman-Weinberg (CW) mechanism \cite{colemanweinberg}. A massless scalar field at tree-level may gain mass due to loop corrections.
In fact, the radiative corrections of a gauge field can provide masses not only for the scalar degree of freedom but also for the gauge mediator itself. 
Indeed, depending on the parameter space of the theory, spontaneous symmetry breaking can take place. The
scalar field acquires a vacuum expectation value (vev) and the
gauge bosons acquire mass through a Higgs-like mechanism.

The CW effective potential that triggers the non-trivial vev is provided by the tree-level contribution from the scalar field quartic self-interaction plus the radiative corrections. Furthermore, within the CW mechanism those introduced by the gauge interactions are assumed to dominate over other possible loop contributions (e.g., loops with scalar self-interactions). 
These corrections can be computed perturbatively for any number of external scalar fields, being dominated by one-loop Feynman diagrams. 

In this work, first, we will briefly review the analysis done by Coleman and Weinberg on quartic self-interaction theories of a charged scalar field interacting with electrodynamics, and second, we will generalize this approach for the gravitational interaction. For such a purpose,  we will assume a particular model for the gravitational mediator:  {\it graviton}. In the first place, we will define the fundamental elements of the theory: Lagrangian density, graviton propagator or interaction vertices. We will study which diagrams contribute to the potential and we will use the aforementioned elements to obtain the potential through the CW mechanism.

In Sec.~\ref{sec:SQED-CW}, we will review the basic example discussed by Coleman and Weinberg, Scalar Quantum Electrodynamics (SQED)~\cite{colemanweinberg}. Starting from a conformal $\lambda \phi^4$ potential, the gauge-boson loops trigger a dynamical symmetry breaking. Sec.~\ref{sec:grav-CW} provides the central results of this article. Therein we study a conformal classical Lagrangian including a scalar field $\phi$ with a non-minimal gravitational coupling. We find that graviton loops are also able to trigger a non-zero vev $\langle \phi\rangle \neq 0$ and an effective Planck Mass. Sec.~\ref{sec:grav-CW2}  generalises this model to cases with more than one scalar field. Our final conclusions are provided in Sec.~\ref{sec:conclusions}, whereas
certain technical expressions have been relegated to App.~\ref{app:projectors}.

\section{Dynamical symmetry breaking through gauge boson loops: SQED}
\label{sec:SQED-CW}

The CW approach studies the effective action $\Gamma$ of a quantum theory including a scalar $\phi$.  
We will focus our attention on the non-derivative terms that only contain classical scalar fields $\phi_c$:  
\begin{equation}\label{4}
    \Gamma\, =\, -\, \int \dif^4 x\, V(\phi_c) \, +\, ...
\end{equation}
where the dots stand for terms with derivatives or including other fields apart of $\phi_c$. The effective potential $V$ is thus defined by the non-derivative part of the effective action $\Gamma$. It will be the central object of the CW analysis.

The simplest classically conformal theory one may consider contains just one real scalar $\phi$ with a quartic self-interaction potential: 
\begin{equation}
    \mathcal{L}=\frac{1}{2}(\partial_\mu \phi)^2-\frac{\lambda}{4!} \phi^4 \ \ .
\end{equation}
At lowest order the effective potential $V(\phi_c)$ coincides with the tree-level potential $V_0(\phi_c)=\lambda \phi^4/4!$ with a trivial vacuum $\langle \phi\rangle=0$. At the loop level, Coleman and Weinberg showed that scalar loops are not able to push this vev to a calculable non-trivial value $\langle \phi\rangle\neq 0$. Indeed, the authors noticed that a purely perturbative analysis leads to inconsistencies that need to be cured by the use of renormalization group equations~\cite{colemanweinberg}.   

Nonetheless, these issues can be solved if we incorporate interactions with additional particles. Starting from a scale invariant Lagrangian, Coleman and Weinberg proposed a perturbative framework where the scalar (or scalars) gains a non-vanishing vev $\langle\phi\rangle \neq 0$ due to the loops of these other states~\cite{colemanweinberg}. 
The simplest example is provided by SQED, 
a renormalizable theory with a massless complex scalar $\Phi=(\phi_1+i\phi_2)/\sqrt{2}$ and quartic potential with a $U(1)$ gauge interaction (with $e$ the corresponding charge). The Lagrangian density can be expressed in terms of the two real scalar fields $\phi_1$ and $\phi_2$ in the form:
\begin{eqnarray}
    \mathcal{L}&=&-\frac{1}{4}(F_{\mu\nu})^2+\frac{1}{2}(\partial_\mu\phi_1-eA_\mu\phi_2)^2+\frac{1}{2}(\partial_\mu\phi_2+eA_\mu\phi_1)^2  
-\frac{\lambda}{4!}(\phi_1^2+\phi_2^2)^2 
\, . 
\end{eqnarray}

The effective potential up to one-loop is given in dimensional regularization by,  
\begin{eqnarray}
V(\phi)&=&\frac{\lambda}{4!}\phi^4
+ i\int \frac{\dif^D k}{(2\pi)^D}\sum_{n=1}^\infty \frac{(D-1)}{2n}\left(\frac{e^2\phi^2}{k^2+i\epsilon}\right)^n 
\nonumber\\
    &&       
    +\,\, i\int \frac{\dif^D k}{(2\pi)^D}\sum_{n=1}^\infty \frac{1}{2n}\left(\frac{\frac{1}{2}\lambda\phi^2}{k^2+i\epsilon}\right)^n 
    + i\int \frac{\dif^D k}{(2\pi)^D}\sum_{n=1}^\infty \frac{1}{2n}\left(\frac{\frac{1}{6}\lambda\phi^2}{k^2+i\epsilon}\right)^n\,,  
\end{eqnarray}
with $\phi^2=\phi_1^2 +\phi_2^2$, and $D$, the space-time dimension. After resumming the loops with an arbitrary number of external legs and renormalizing the ultraviolet divergences in the $\overline{\text{MS}}$-scheme one gets the finite potential,   
\begin{eqnarray}\label{eq:Vgauge}
    V(\phi)&=&\frac{\lambda}{4!}\phi^4
     +\frac{3 e^4 \phi^4}{64\pi^2}  \left[\ln\left(e^2\frac{\phi^2}{M^2}\right)  
     -   \frac{5}{6}  \right] 
\nonumber\\
    &&
    +\,\, \frac{\lambda^2 \phi^4}{256\pi^2}\bigg\{ \left[\ln\left(\frac{\lambda}{2}\frac{\phi^2}{M^2}\right)-\frac{3}{2}\right]   
    +\frac{1}{9}\left[\ln\left(\frac{\lambda}{6}\frac{\phi^2}{M^2}\right)-\frac{3}{2}\right] \bigg\} \,, 
\end{eqnarray}
where $M$ stands for the renormalization scale. For simplicity, the computation is performed in the Landau gauge~\cite{colemanweinberg}. Perturbation theory requires that the $\mO(\lambda^2)$ terms are subdominant with respect to the tree-level potential, which is $\mO(\lambda)$. Hence, these loops cannot be responsible for bending the leading order (LO) scalar potential to generate a non-trivial minimum.     
However, perturbativity does not imply that the $\mO(e^4)$ terms need to be suppressed with respect to the $\mO(\lambda)$ one. Thus, 
Coleman and Weinberg neglects these $\mO(\lambda^2)$ scalar boson loops, 
so the effective potential is essentially provided by the $\mO(\lambda)$ tree-level contribution and the $\mO(e^4)$ gauge boson loops.  
Under the assumption $\lambda^2\ll |\lambda|\sim  e^4$, the model is now able to trigger the spontaneous symmetry breaking and generate a non-trivial minimum given by the condition,        
\begin{equation}
    \frac{\dif V}{\dif \phi}\bigg|_{\langle \phi \rangle}=0 \qquad \implies \qquad \lambda=\frac{9}{8\pi^2}e^4\left[\frac{1}{3}  
    -\ln\left(e^2\frac{\langle\phi\rangle^2}{M^2}\right)\right] \,,
\label{eq:SQED-minimum}
\end{equation}  
with $\lambda$ and $e$ renormalized at the scale $M$. 
Note that perturbativity is still valid around the minimum extracted above  as far as $e^4/(4\pi)^2 \sim |\lambda|  \ll \lambda^2/(4\pi)^2$.  
Substituting this vev condition in the potential leads to the final expression,  
\begin{equation}\label{34}
    V(\phi)=\frac{3e^4}{64\pi^2}\phi^4\left[\ln \left(\frac{\phi^2}{\langle \phi \rangle ^2}\right)-\frac{1}{2}\right] \, ,
\end{equation}
where we have changed the dependency on $\lambda$ by that on  $\langle\phi\rangle$ by means of~(\ref{eq:SQED-minimum}), this is, we have changed a dimensionless parameter by a dimensional one. This phenomenon is known as dimensional transmutation~\cite{colemanweinberg} .

The spontaneous symmetry breaking generates one massless boson~\cite{goldstone} and a real scalar $S=\phi-\langle\phi\rangle$, with mass,  
\begin{equation}
    m_S^2=\frac{\dif ^2V}{\dif \phi^2}\Bigg|_{\langle \phi \rangle}=\frac{3e^4}{8\pi^2}\langle \phi \rangle ^2 \, .
\end{equation}
Due to the gauge nature of the $U(1)$ symmetry, the massless boson mixes with the $U(1)$ gauge boson and leads to a physical vector boson $V$ with mass~\cite{colemanweinberg}, 
\begin{equation}
    m_V^2\, =\, e^2\langle \phi \rangle ^2 \ \ .
\end{equation}

\section{Gravitational interaction}
\label{sec:grav-CW}

Next, we shall study the CW mechanism for the gravitational interaction. For this purpose, we should introduce a gravitational theory without explicit dimensional parameters. There have been numerous proposals about scale-free gravitational interactions. For instance, the idea of conformal gravity has been recurrent from different approaches~\cite{Fradkin:1978yf, Zee:1983mj, Shapiro:1994st, Hamada:2002cm, Hamada:2009hb, Donoghue:2016xnh, Alvarez:2017spt, Donoghue:2018izj}. In the same way that the mass term of a scalar field can be provided by dimensional transmutation, quantum radiative corrections can induce the Planck scale \cite{Adler:1982ri}. For analysing this question, we introduce the Lagrangian density with which we shall work:
\begin{eqnarray}
   \sqrt{|\text{det}\, g|}\, \mathcal{L}_G&=&\sqrt{|\text{det}g|}\left(\frac{R^2}{6f_0^2}+\frac{\frac{1}{3}R^2-R_{\mu\nu}^2}{f_2^2} 
+ \frac{1}{2}(\partial_\mu \phi)^2-\frac{\lambda}{4!} \phi^4-\frac{\xi}{2} \phi^2 R\right) \, . 
\label{eq:L-grav}
\end{eqnarray}
Except for a total derivative, in 3+1 dimensions, this Lagrangian density is the most general one associated to gravity that does not contain dimensional constants. When the scalar field $\phi$ develops a non-trivial vev $\langle\phi\rangle\neq 0$, a reduce Planck mass $\bar{M}_{\rm pl}^2=\xi\langle\phi\rangle^2$ is generated in the Lagrangian. Likewise, the physical spectrum of the above action contains the usual massless spin-two graviton, a massive spin-two mediator with mass $m_2^2=f_2^2\bar{M}_{\rm pl}^2/2$, and a massive scalar mode with mass $m_0^2=f_0^2\bar{M}_{\rm pl}^2/2$~\cite{Stelle:1976gc, Stelle:1977ry} in the limit when it does not mix with $\phi$ 
\footnote{In general, the mixing between the gravitational scalar mode and $\phi$ cannot be neglected as we discuss in App.~\ref{app:Masses_SS} and it has been pointed out in Refs. \cite{agravity,Kubo:2022dlx}.}.
The massive scalar graviton has an interesting phenomenology and it has been proposed as inflaton \cite{Starobinsky:1980te} or dark matter \cite{Cembranos:2008gj}. The role of the massive spin-two graviton is more controversial due to the fact that its kinetic term is ghost-like and it potentially leads to unitarity violations.
Recent works claim that this problem could be solved, for instance, 
with an alternative definition of probability~\cite{Salvio:2019ewf,Salvio:2020axm}.   
Furthermore, within the standard probability approach, negative-energy ghosts can be well-defined in classical mechanics, quantum mechanics and classical field theory~\cite{Gross:2020tph}. However, there is not a consensus about this serious issue that in addition to the unitarity problem could introduce breaking of causality and inadmissible instabilities \cite{Simon:1991bm}. On the other hand, the model is renormalizable, what makes it particularly interesting for the analysis we develop within this work. Indeed,
this type of gravitational theories have been studied in different contexts \cite{Stelle:1976gc, Stelle:1977ry,agravity, Karam:2018mft, Gialamas:2020snr, Kubo:2020fdd, Gialamas:2021enw}. We will follow a similar approach to that followed within the {\it agravity} model \cite{agravity}. 

The real scalar field $\phi$ is coupled to gravity through different interaction terms. Again, there is no tree-level mass term for the scalar field. On the one hand, we have the kinetic part of the graviton described by the following elements: $R$ is the curvature scalar, $R_{\mu\nu}$ is the Ricci tensor and $f_0$ and $f_2$ are the dimensionless constants that are related to the gravitational coupling. On the other hand, we have the kinetic and potential terms of the scalar field that we have already studied previously. Finally, the interaction between the scalar field and the gravitons depends also on the coupling constant $\xi$, which is also dimensionless.

To continue with the study of the interaction, we expand the metric tensor as a perturbation with respect to the Minkowski geometry as follows:
\begin{equation}
    g_{\mu\nu}=\eta_{\mu\nu}+\kappa h_{\mu\nu} \ \ .
\end{equation}
where $\eta_{\mu\nu}$ is the Minkowski spacetime metric, and $h_{\mu\nu}$ is the perturbation describing the graviton. At this point, $\kappa$ is an arbitrary parameter which allows to normalize canonically the perturbation $h_{\mu\nu}$ and its propagator. We can write this propagator as: 
\begin{equation}
    \hspace{-2em}\vcenter{\includegraphics[width=0.2\textwidth]{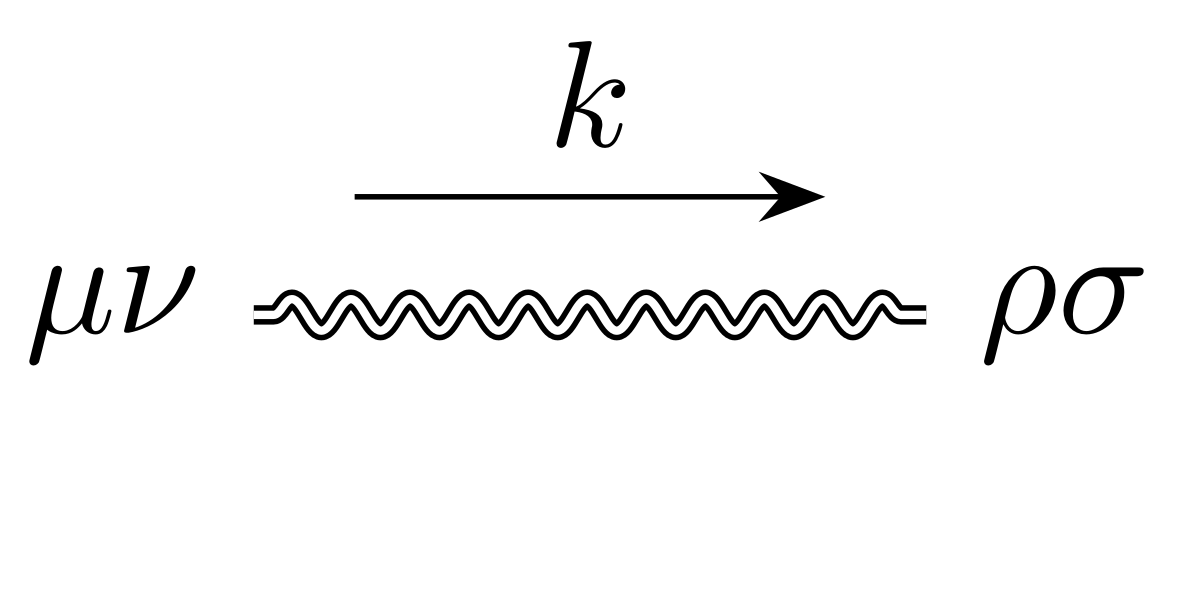}}\hspace{-29.5em}
    \equiv D_{\mu\nu\rho\sigma}= \frac{1}{\kappa^2}\frac{i}{k^4}
    \left[-2f_2^2P^{(2)}_{\mu\nu\rho\sigma}+f_0^2P^{(0)}_{\mu\nu\rho\sigma}+2 \zeta_{\rm gf} \left(P^{(1)}_{\mu\nu\rho\sigma}+\frac{P^{(0w)}_{\mu\nu\rho\sigma}}{2}\right)\right]\, , 
\end{equation}
with $\zeta_{\rm gf}$, the gauge-fixing parameter.
The propagator is expressed by means of projectors on the components of the graviton of spin 2, 1 and 0. It is worth noting the presence of the \textit{gauge fixing} term.  
The expressions of these projectors are the following \cite{agravity}:
\begin{eqnarray}
   && P^{(2)}_{\mu\nu\rho\sigma}=\frac{1}{2}T_{\mu\rho}T_{\nu\sigma}+\frac{1}{2}T_{\mu\sigma}T_{\nu\rho}-\frac{1}{D-1}T_{\mu\nu}T_{\rho\sigma} \ \ ,
\nonumber\\
    &&    
    P^{(1)}_{\mu\nu\rho\sigma}=\frac{1}{2}\left(T_{\mu\rho}L_{\nu\sigma}+T_{\mu\sigma}L_{\nu\rho}+T_{\nu\rho}L_{\mu\sigma}+T_{\nu\sigma}L_{\mu\rho}\right) \ \ ,
\nonumber\\
    &&
    P^{(0)}_{\mu\nu\rho\sigma}=\frac{1}{D-1}T_{\mu\nu}T_{\rho\sigma} \ \ ,
\nonumber\\
    &&
    P^{(0w)}_{\mu\nu\rho\sigma}=L_{\mu\nu}L_{\rho\sigma} \,,
\end{eqnarray}
where
\begin{equation}
    T_{\mu\nu}=\eta_{\mu\nu}-\frac{k_\mu k_\nu}{k^2} \ \ , \ \ L_{\mu\nu}=\frac{k_\mu k_\nu}{k^2} \ \ , \ \ D=4-2\epsilon \,.
\end{equation}

\begin{figure}[!t]
    \centering
    \includegraphics[width=0.7\textwidth]{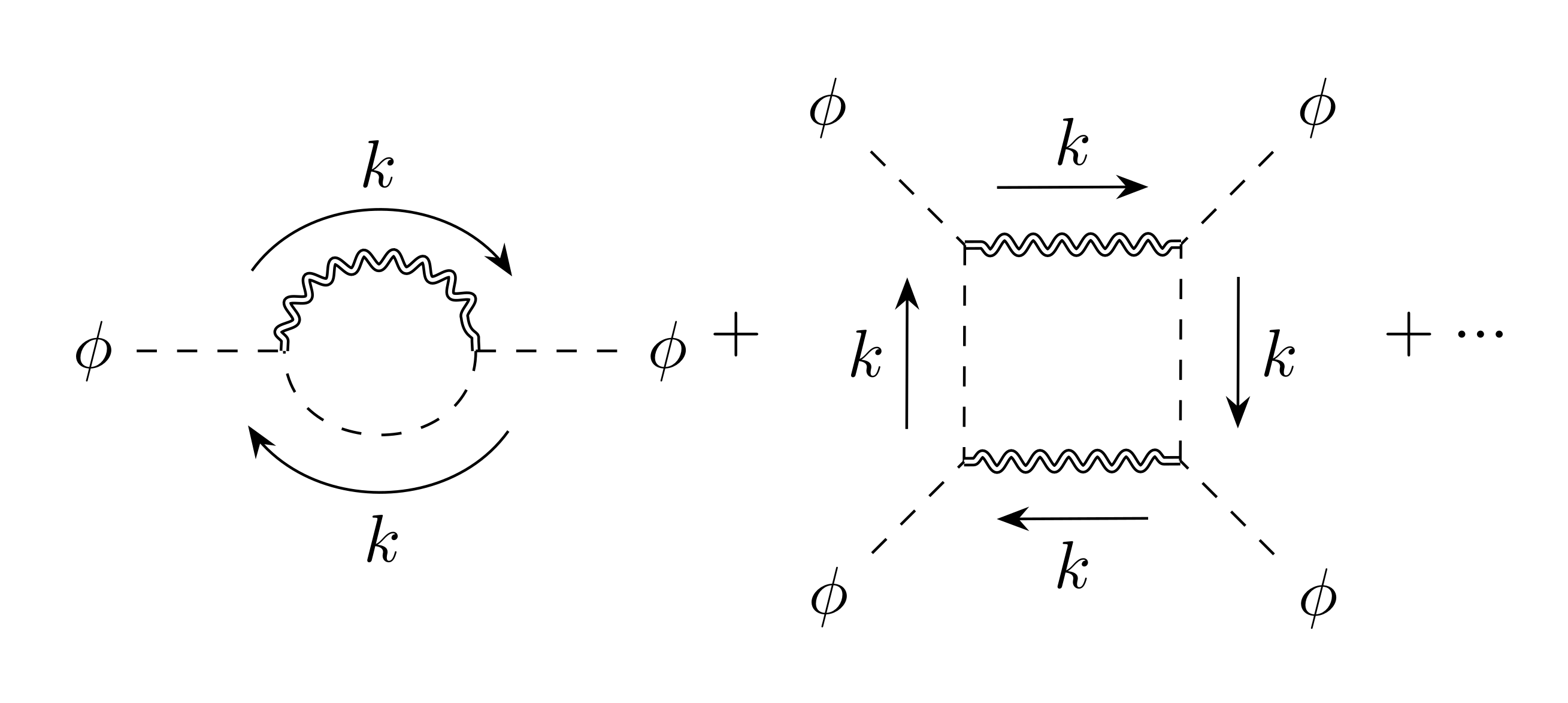}
    \caption{1PI diagrams of the $\phi^2h$ interaction contributing to the potential at the one-loop approximation.}
    \label{F6}
    \label{fig:mixed-G-diagrams}
\end{figure}

To find the vertex of two scalar fields with one and two gravitons we must expand $\sqrt{|\text{det}g|}\,R$ to the linear and quadratic order of the perturbation $h_{\mu\nu}$, respectively. To carry out the perturbative expansion we must study how this perturbation affects the geometrical quantities:
\begin{equation}
    \Gamma^\rho_{\mu\nu}=\frac{\kappa}{2}\left(\eta^{\rho\sigma}-\kappa h^{\rho\sigma} \right)\left(\partial_\mu h_{\sigma\nu}+\partial_\nu h_{\mu\sigma}-\partial_\sigma h_{\nu\mu}\right)+\mathcal{O}(h^3) \ \ ,
\end{equation} 
\begin{equation}
    R = \left(\eta^{\mu\rho}-\kappa h^{\mu\rho}\right)\left(\partial_\nu\Gamma^\nu_{\mu\rho}-\partial_\mu\Gamma^\nu_{\nu\rho} + 
    \Gamma^\lambda_{\mu\rho}\Gamma^\nu_{\lambda\nu}-\Gamma^\lambda_{\nu\rho}\Gamma^\nu_{\lambda\mu}\right)+\mathcal{O}(h^3) \ \ ,
\end{equation}            
where we have used that \cite{weakfield}
\begin{equation}
    g^{\mu\nu}=\eta^{\mu\nu}-\kappa h^{\mu\nu}+\mathcal{O}(h^2) \ \ .
\end{equation}
Taking into account the above results, we can write the following perturbative equality: 
\begin{eqnarray}
&&    \sqrt{|\text{det}g|}\,R = 
    \kappa \left(\partial_\mu\partial_\nu h^{\mu\nu} -\partial^2 h\right) +\kappa^2\left(h^{\mu\nu}\partial_\mu\partial_\nu h
    +h^{\mu\nu}\partial^2h_{\mu\nu}\right.
\nonumber\\
&&\;\;
    -\,\, 2h^{\mu\nu}\partial_\mu\partial_\rho h^\rho_\nu 
    -\partial_\mu h^{\mu\nu}\partial_\rho h^\rho_\nu + \partial_\mu h^{\mu\nu}\partial_\nu h+ \frac{3}{4}\partial_\rho h_{\mu\nu}\partial^\rho h^{\mu\nu}
\nonumber\\
&&\;\;
    -\left.\frac{1}{4}\partial_\mu h \partial^\mu h-\frac{1}{2}\partial_\rho h_{\mu\nu}\partial^\mu h^{\nu \rho}
    +\frac{1}{2}h\partial_\mu\partial_\nu h^{\mu\nu}-\frac{1}{2}h\partial^2 h\right)+\mathcal{O}(h^3) \ \ ,
\end{eqnarray}            
where $h\equiv h_\mu^\mu$ and we have used that \cite{weakfield}
\begin{equation}
    \sqrt{|\text{det}g|}=1+\frac{\kappa}{2}h +\mathcal{O}(h^2) \, .
\end{equation}
Therefore, we can write the following expressions for the vertices:
\begin{equation}
\hspace{-2em}\vcenter{\includegraphics[width=0.16\textwidth]{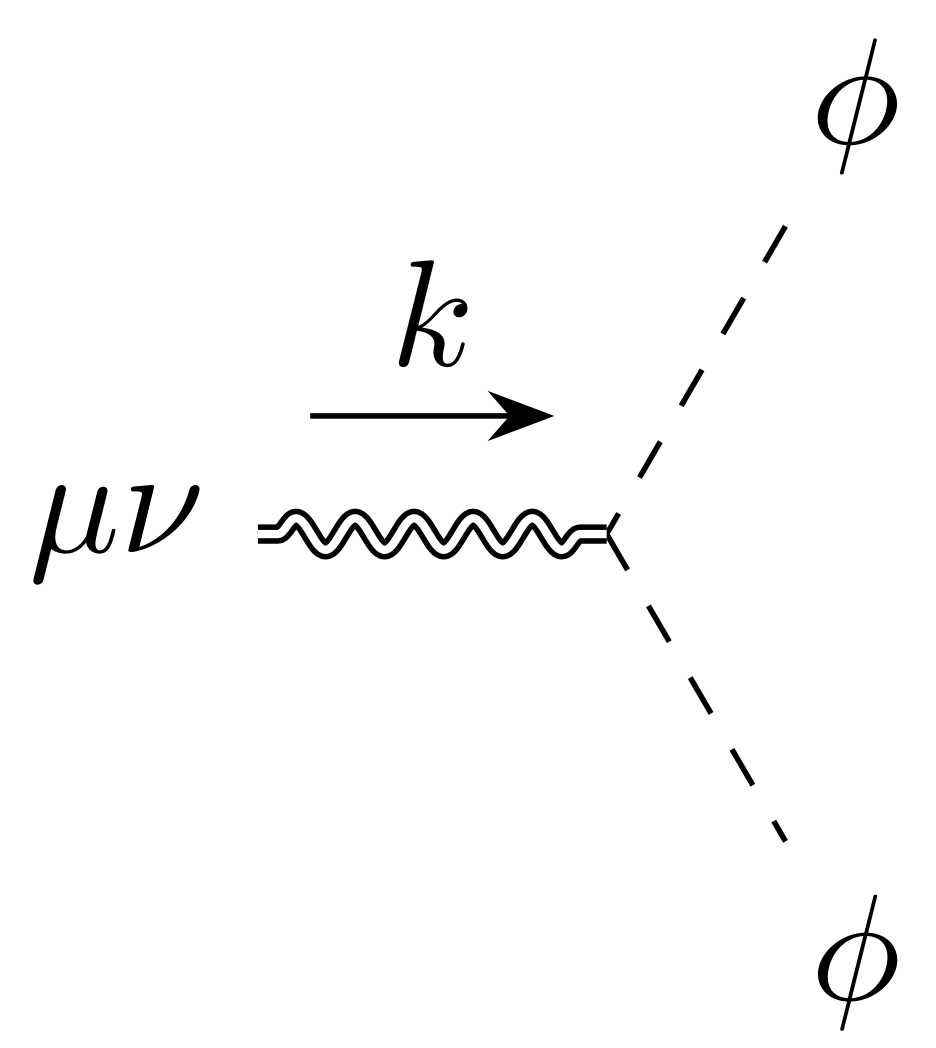}}\hspace{-30.5em}
\equiv C_{(1\xi),\mu\nu}= i \kappa \xi \left(k_\mu k_\nu -k^2\eta_{\mu\nu}\right)  = 
-i\kappa \xi \, k^2 T_{\mu\nu}
\,,
\end{equation}
\begin{eqnarray}
&&
    \hspace{-1em}\vcenter{\includegraphics[width=0.19\textwidth]{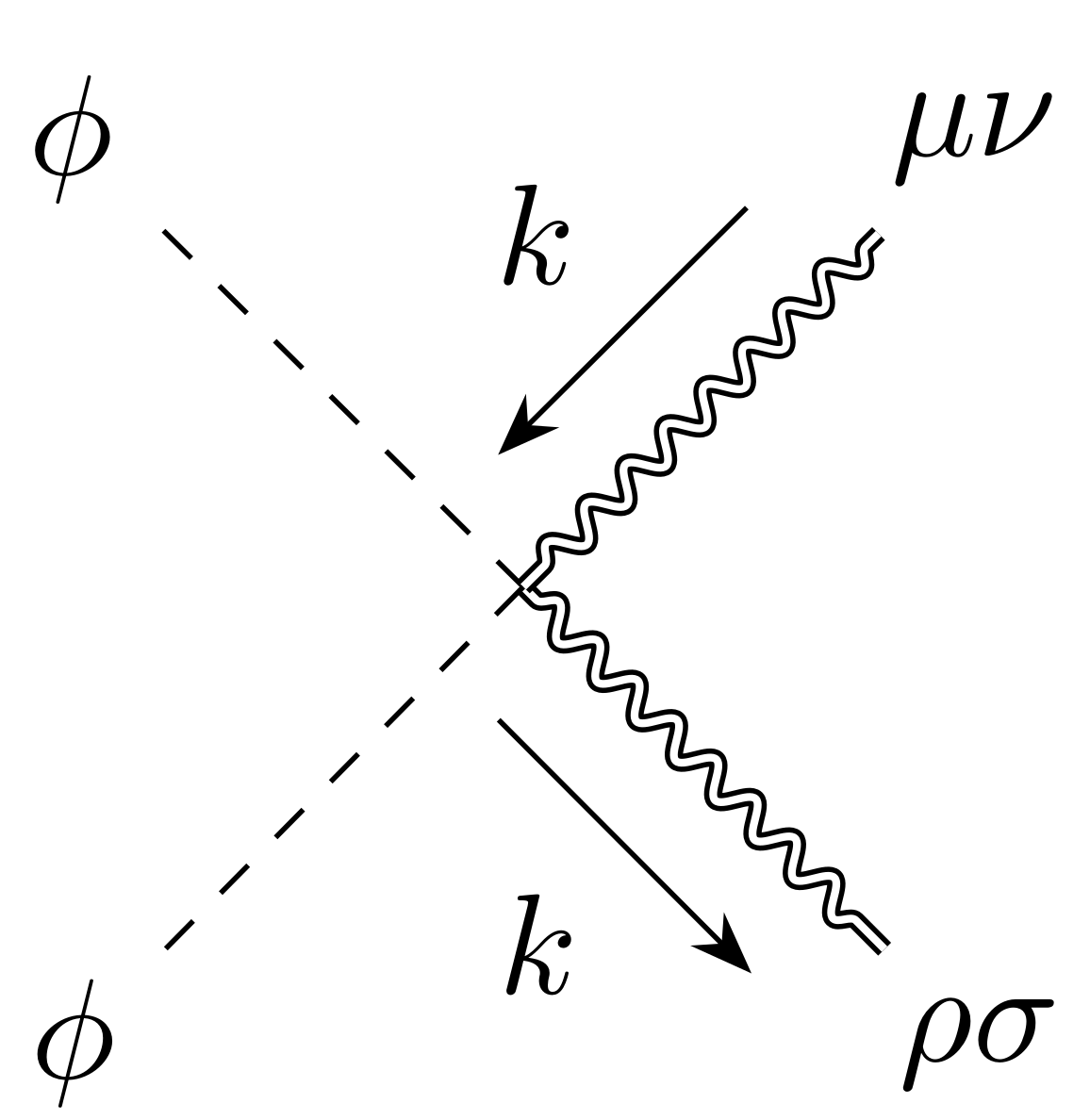}}\hspace{-30em}
    \equiv C_{(2\xi),\mu\nu\rho\sigma}=i\kappa^2\xi \bigg[A_{\mu\nu\rho\sigma}+A_{\rho\sigma\mu\nu} 
\nonumber\\
&&\;\;\;\;\;\;\;\;\;\;\;\;\;\;\;\;\;
-\,\, \frac{1}{2}\left(A_{\nu\sigma\mu \rho}+A_{\mu\sigma\nu \rho}+A_{\nu\rho\mu \sigma}+A_{\mu\rho\nu\sigma}\right) 
+k^2\left(\frac{1}{2}B_{\mu\rho\nu\sigma}+\frac{1}{2}B_{\mu\sigma\nu\rho}-B_{\mu\nu\rho\sigma}\right)\bigg] 
\nonumber\\
&& \hspace*{2.5cm} =i\kappa^2 \xi \, k^2 \left(P^{(2)}_{\mu\nu\rho\sigma} - (D-2) P^{(0)}_{\mu\nu\rho\sigma}\right) \,,
\end{eqnarray}
where $A_{\mu\nu\rho\sigma}=k_\mu k_\nu \eta_{\rho\sigma}$ and $B_{\mu\nu\rho\sigma}=\eta_{\mu\nu }\eta_{\rho\sigma}$. We could continue with the expansion to order $h^3$ to obtain the expression for the interaction vertex of two scalar fields with three gravitons. However, they will be only relevant for diagrams that contribute to the potential at two or higher loops, which would depart from the one-loop leading order approach we are computing. In this theory, unlike what happened in electrodynamics, the $h\phi^2$ interaction does contribute. Therefore, we also need to study the diagrams represented in Figure \ref{F6}.

\begin{figure}[!t]
    \centering
    \includegraphics[width=1\textwidth]{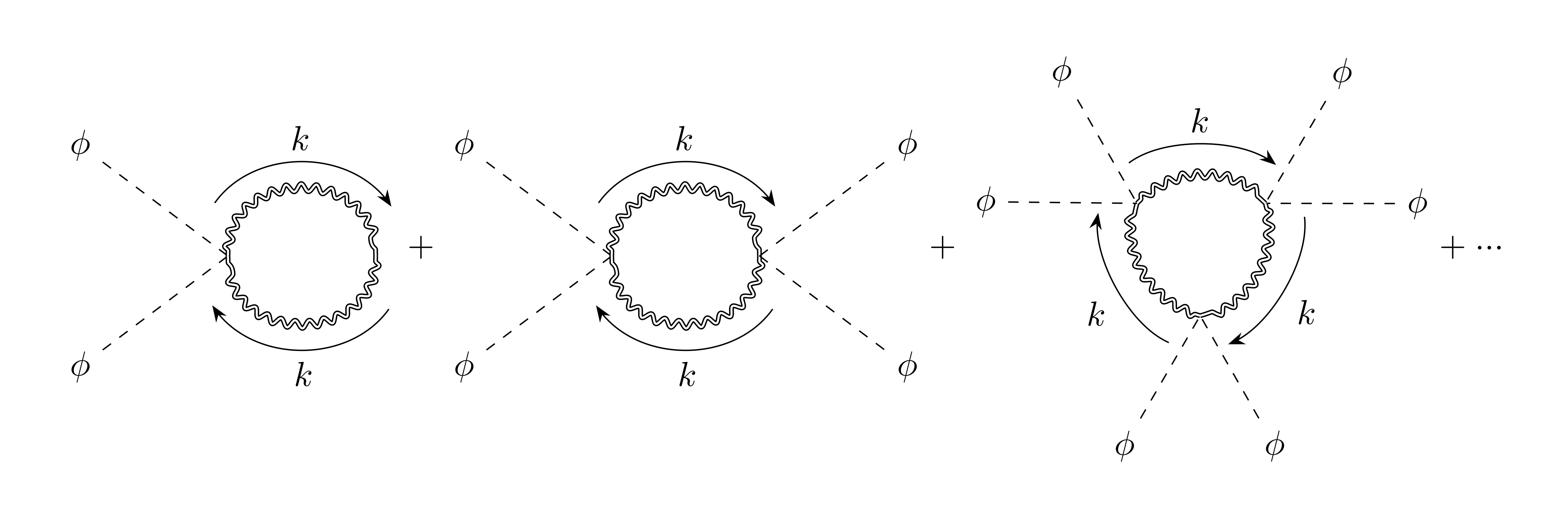}
    \caption{1PI diagrams of the interaction $\phi^2h^2$ that contribute to the potential in a loop approximation.}
    \label{F7}
    \label{fig:pure-G-diagrams}
\end{figure}

\begin{figure}
    \centering
    \includegraphics[width=0.7\textwidth]{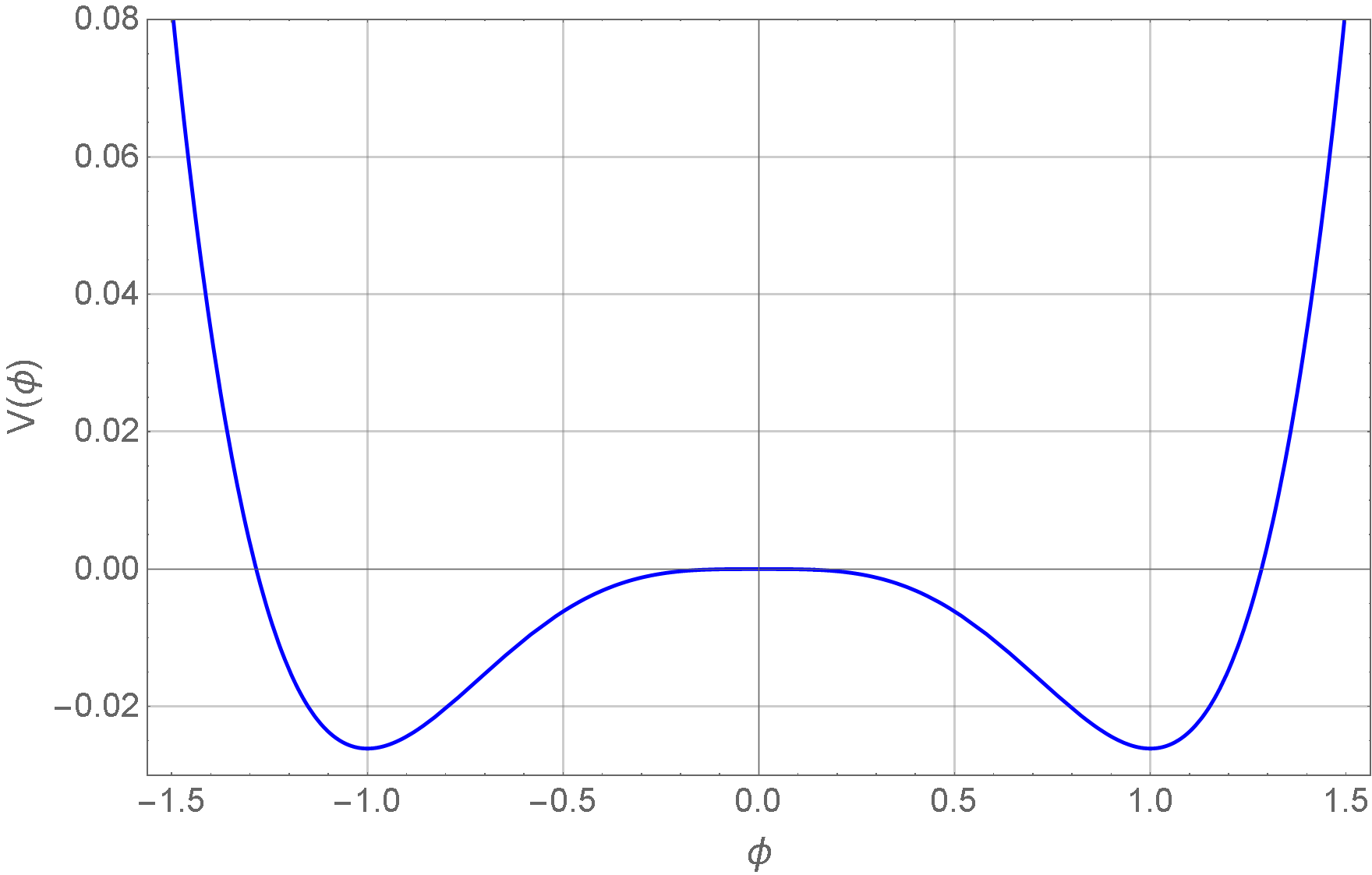}
    \caption{CW Potential in Eq.~\eqref{eq:V-grav2} as a function of the scalar field for $\xi=f_0=f_2=\langle\phi\rangle=1$.}
    \label{F8}
\end{figure}

When performing the contraction of the vertices and the propagators in this type of diagrams, we see that only the projector term $P^{(0)}$ contributes. This is the result for $2n$ vertices:
\begin{equation}
   \left( \frac{i}{k^2}\, C^{\mu\nu}_{(1\xi)}\, D_{\mu\nu\rho\sigma}\, C^{\rho\sigma}_{(1\xi)}   
    \right)^n   =\left(\frac{(D-1)f_0^2\xi^2}{k^2}\right)^n \,.  
\label{eq:mixed-grav-diag}  
\end{equation}
On the other hand, we have the interaction $\phi^2h^2$ to which the diagrams presented in Figure \ref{F7} correspond. The contraction of this type of diagrams is non-zero for the projectors $P^{(0)} $ and $P^{(2)}$. For $n$ vertices:
\begin{equation} 
{\rm Tr}\left\{ \left(D\cdot C_{(2\xi)}\right)^n \right\}     
    =\left(\frac{2\xi}{k^2}\right)^n\left[\left(\frac{(D-2)}{2} f_0^2\right)^n\, +\, \frac{(D^2-D-2)}{2}  f_2^{2n} \right] \, , 
\label{eq:pure-grav-diag}
\end{equation}
with the trace Tr$C\equiv C^{\mu\nu}_{\,\,\, \mu\nu}$ and the dot-product defined as $(A\cdot B)^{\mu\nu\, \rho\sigma}\equiv A^{\mu\nu}_{\,\,\, \alpha\beta} B^{\alpha\beta\, \rho\sigma}$ for the tensors $A$ and $B$.
For the results in Eqs.~(\ref{eq:mixed-grav-diag}) and (\ref{eq:pure-grav-diag}), we have made use of the projector relations summarised in App.~\ref{app:projectors}. 
Note that the gauge fixing term of the propagator vanishes when contracted with the vertices. This is of great importance since it implies that the types of diagrams that contribute to the potential are independent of the gauge, at least at one-loop order. In short, we can now write the expression we obtain for the potential:
\begin{eqnarray}
    V(\phi)&=&\frac{\lambda}{4!}\phi^4   
+ i\int \frac{\dif^4k}{(2\pi)^4}\sum_n \frac{1}{2n}\left(\frac{(D-1)  f_0^2\xi^2 \phi^2}{k^2+i\epsilon}\right)^n
\nonumber\\
&& 
\hspace*{-1cm} +\,\, i\int \frac{\dif^4k}{(2\pi)^4}\sum_n \frac{1}{2n}\left(\frac{(D-2)  f_0^2\xi \phi^2}{k^2+i\epsilon}\right)^n 
+ i\, \frac{(D^2-D-2)}{2}\,\int \frac{\dif^4k}{(2\pi)^4}\sum_n \frac{1}{2n}\left(\frac{2f_2^2 \xi \phi^2}{k^2+i\epsilon}\right)^n 
\nonumber\\
&& 
\hspace*{-0.5cm} +\,\, i\int \frac{\dif^4k}{(2\pi)^4}\sum_n \frac{1}{2n}\left(\frac{\frac{1}{2}\lambda \phi^2}{k^2+i\epsilon}\right)^n \, .
\end{eqnarray} 
Resumming the contributions for any number of external legs and renormalizing the ultraviolet divergences in the $\overline{\text{MS}}$-scheme leads to the effective potential,  
\begin{eqnarray}
    V(\phi)&=& \frac{\lambda}{4!}\phi^4 
    +   \frac{   9\xi^4f_0^4\phi^4 }{64\pi^2} \left[\ln\left(3\xi^2f_0^2\frac{\phi^2}{M^2}\right)-\frac{3}{2}\right]
    +   \frac{  \xi^2f_0^4 \phi^4  }{16\pi^2}\left[\ln\left(2\xi f_0^2\frac{\phi^2}{M^2}\right)-\frac{3}{2}\right]
\nonumber\\
&&
    +\,\, \frac{  5\xi^2f_2^4 \phi^4}{16\pi^2}\left[\ln\left(2\xi f_2^2\frac{\phi^2}{M^2}\right)-\frac{1}{5} \right]
    +\frac{\lambda^2 \phi^4}{256\pi^2} \left[\ln\left(\frac{\lambda}{2}\frac{\phi^2}{M^2}\right)-\frac{3}{2}\right]  \, .
\label{eq:V-grav1}
\end{eqnarray} 
As it happens in the basic example of Coleman and Weinberg, the scalars loops can never overcome the tree-level potential and generate a non-trivial minimum, as $\mO(\lambda^2)$ corrections are subdominant with respect to the tree-level $\mO(\lambda)$ term. However, this is not the case with the loops of the remaining particles of the theory. Indeed, we shall see that the gravity loops deform the effective potential and trigger the spontaneous symmetry breaking. 
Following Coleman and Weinberg construction, we will consider $\xi f_j^2/(4\pi)^2 \sim \xi^2 f_j^2/(4\pi)^2 \sim |\lambda| \gg \lambda^2/(4\pi)^2 $. Under this hypothesis, scalar loops associated with $\mO(\lambda^2)$ contributions, can be then neglected for simplicity of the derivation that follows. 
$V(\phi)$ has now a minimum $\langle \phi\rangle\neq 0$ provided by the relation:
\begin{eqnarray} 
\left.\frac{\dif V}{\dif \phi}\right|_{\phi=\langle\phi\rangle}=0  \,,
\end{eqnarray}
which implies,  
\begin{eqnarray}
\lambda&=&\frac{3}{8\pi^2}\left\{9\xi^4f_0^4\left[1-\ln\left(3\xi^2f_0^2\frac{\langle\phi\rangle^2}{M^2}\right)\right]\right.
\nonumber
\\
&&  
+\,\, 4\xi^2f_0^4\left[1-\ln\left(2\xi f_0^2\frac{\langle\phi\rangle^2}{M^2}\right)\right]
+\left.20\xi^2f_2^4\left[\frac{3}{10}-\ln\left(2\xi f_2^2\frac{\langle\phi\rangle^2}{M^2}\right)\right]\right\} \, .
\label{eq:1S-minimum-rel}
\end{eqnarray}
This equality establishes a relationship between $\lambda$ and the scalar field vev. So we can eliminate that coupling from the effective potential and express it in terms of $\langle \phi\rangle$ and the gravitational couplings:   
\begin{equation}
    V(\phi)=\frac{\xi^2}{64\pi^2}\left(9f_0^4\xi^2+4f_0^4+20f_2^4\right)\phi^4 \left[\ln\left(\frac{\phi^2}{\langle \phi\rangle^2}\right)-\frac{1}{2}\right] \ \ .  
\label{eq:V-grav2}
\end{equation}
This potential is represented in Fig.~\ref{F8}.   
Notice that the potential contains terms of order $\xi^2$ and $\xi^4$. In principle, we are implicitly assuming $\xi\sim \mO(1)$ and all contributions are kept in our results. Nonetheless, $V(\phi)$ can be further simplified in the case of models with either $\xi\ll 1$ or $\xi\gg 1$.

Expanding the potential around the vacuum state, one can extract the mass of the physical scalar field $S=\phi- \langle \phi\rangle$, which now becomes massive: 
\begin{equation}
    m_S^2=\left.\frac{\dif^2V}{\dif \phi^2}\right|_{\phi=\langle \phi\rangle}=\frac{\langle\phi\rangle^2}{8\pi^2}\xi^2\left(9f_0^4\xi^2+4f_0^4+20f_2^4\right) \,.
\end{equation}
Furthermore, as $\phi^2$ gets a non-vanishing vev $\langle\phi\rangle^2\neq 0$, the Lagrangian~(\ref{eq:L-grav}) gets an effective term proportional to $R$. This generates the Planck scale and leads to the following relations between the latter and the scalar mass:
\begin{equation}
\xi\langle\phi\rangle^2= \bar{M}^2_{\text{Pl}} 
\quad \implies \quad     m_S^2=   
\frac{\bar{M}_{\text{Pl}}^2}{8\pi^2}
\xi\left(9f_0^4\xi^2+4f_0^4+20f_2^4\right) \ \ , 
\end{equation}
where $\bar{M }_{\text{Pl}}$ is the reduced Planck mass \cite{agravity,masaplanck}. 

\section{Gravitational interaction with two scalar fields}
\label{sec:grav-CW2}

It is also interesting to extend the study to the case of several scalar fields. In this section, we will discuss how the mechanism behaves if we have two real scalar fields $\phi_1$ and $\phi_2$. They couple to each other and to gravity through the following renormalizable Lagrangian:
\begin{eqnarray}
\sqrt{|\text{det}g|}\,\mathcal{L}&=&\sqrt{|\text{det}g|}\bigg[\frac{1}{2}(\partial_\mu \phi_1)^2  +\frac{1}{2}(\partial_\mu \phi_2)^2
- V_0(\phi_1,\phi_2)
\nonumber\\
&&    
\hspace*{1.5cm}  
+\,\, \frac{R^2}{6f_0^2}+\frac{\frac{1}{3}R^2-R_{\mu\nu}^2}{f_2^2} 
-\frac{1}{2}R(\xi_1\phi_1^2+\xi_2\phi_2^2)\bigg] \,,
\label{59}
\end{eqnarray}
with the general tree-level scalar potential, 
\begin{eqnarray}
V_0(\phi_1,\phi_2) &=&   \frac{1}{4!}\left(\lambda_{11}\phi_1^4+\lambda_{22}\phi_2^4+2\lambda_{12} \phi_1^2 \phi_2^2 \right) \,= \, 
\frac{\lambda}{4!}\phi^4 + \frac{\alpha}{4!} \phi^2 (\phi_1^2-\phi_2^2) 
+\frac{\beta}{4!} (\phi_1^2-\phi_2^2)^2 \, , 
\nonumber\\
\end{eqnarray}
with $\phi^2=\phi_1^2+\phi_2^2$. 
In the general case, there are two different non-minimal gravitational couplings: $\xi_1$ and $\xi_2$; and three independent quartic couplings of the scalar sector: $\lambda_{11},\, \lambda_{22},\, \lambda_{12}$.  
For convenience, it may be useful to express the latter quartic couplings in terms of $\lambda=(\lambda_{11}+\lambda_{22}+2\lambda_{12})/4$,  
$\alpha=(\lambda_{11}-\lambda_{22})/2$ and 
$\beta=(\lambda_{11}+\lambda_{22}-2\lambda_{12})/4$. 

We can derive the one-loop potential following the previous discussion for one real scalar. On the one hand, for the diagrams with gravitational interactions of the type $\phi^2h$, where we had $\xi^2 \phi^2$ for the single scalar case, now we have $\xi_1^2 \phi_2^2+\xi_2^2 \phi_2^ 2$. On the other hand, for interactions of the type $\phi^2h^2$, we substitute $\xi\phi^2$ for $\xi_1\phi_1^2+\xi_2\phi_2^2$. 
Thus, the one-loop correction to the effective potential due to gravitational loops (Figs.~\ref{fig:mixed-G-diagrams} and \ref{fig:pure-G-diagrams}) has the following form:  
\begin{eqnarray}
V(\phi_1,\phi_2)^{\rm 1\ell oop}
&=&
\frac{9f_0^4}{64\pi^2}\left(\tilde{\phi}^T\tilde{\xi}^2\tilde{\phi}\right)^2\left[\ln\left(3f_0^2\frac{\tilde{\phi}^T\tilde{\xi}^2\tilde{\phi}}{M^2}\right)-\frac{3}{2}\right]
\\
&& \hspace*{-1cm} 
+\,\, \frac{1}{16\pi^2}\left(\tilde{\phi}^T\tilde{\xi}\tilde{\phi}\right)^2\left\{f_0^4\left[\ln\left(2 f_0^2\frac{\tilde{\phi}^T\tilde{\xi}\tilde{\phi}}{M^2}\right)-\frac{3}{2}\right]+5f_2^4\left[\ln\left(2 f_2^2\frac{\tilde{\phi}^T\tilde{\xi}\tilde{\phi}}{M^2}\right)
-\frac{1}{5}\right]\right\}  \, ,
\nonumber 
\end{eqnarray}
where we have defined:
\begin{equation}
    \tilde{\phi}=
    \begin{pmatrix}
        \phi_1 \\
        \phi_2
    \end{pmatrix} \ \ , \ \ \ \ \ \ 
    \tilde{\xi}=
    \begin{pmatrix}
        \xi_1 & 0 \\
        0 & \xi_2 
    \end{pmatrix} 
     = \xi \, \left[ \mbox{\bf 1} +\hat{\delta} \begin{pmatrix}
        1 & 0 \\
        0 & -1 
    \end{pmatrix} \right]\, ,
\end{equation}
with $\xi_{1,2}\equiv \xi\pm\delta \equiv \xi(1\pm \hat{\delta})$.
In this section, we define $\xi$ as the average of the two non-minimal couplings: $\xi=(\xi_{1}+\xi_{2})/2$;
whereas $\delta$ is their difference divided by two: $\delta=(\xi_{1}-\xi_{2})/2$. $\hat{\delta}$ accounts for the same difference but normalized to the average value: $\hat{\delta}=\delta/\xi$.
Following the CW approach, we will assume that these contributions are of the same order as the tree-level potential and, hence, scalar boson loops are subdominant and can be neglected.

\subsection{$U(1)$ symmetry limit}

It is interesting to pay particular attention to the case in which the two scalars are characterised by similar self-interactions and to the scalar curvature. In theories with $\xi_1\simeq \xi \simeq \xi_2$ and $\lambda_{11}\simeq \lambda_{22}\simeq \lambda_{12}$, there is an approximated global $U(1)$ symmetry and a Nambu-Goldstone boson shows up if it is spontaneously broken. It is convenient to decompose the real scalar doublet in polar coordinates in the form $\phi_1=\phi\cos\theta$, $\phi_2=\phi\sin\theta$, where $\theta$ will be related to the mentioned Nambu-Goldstone mode.

In the $U(1)$ symmetric limit $\xi_1=\xi_2=\xi$ and $\lambda_{11}= \lambda_{22}= \lambda_{12}=\lambda$, the CW potential becomes:
\begin{eqnarray}
V(\phi,\theta) &=& 
\frac{\lambda}{4!}\phi^4
+\frac{9\xi^4f_0^4\phi^4}{64\pi^2} 
  \left[\ln\left(\frac{3\xi^2f_0^2 \phi^2}{M^2}\right)-\frac{3}{2}\right]
\nonumber\\
&&
\quad        +\,\, \frac{\xi^2\phi^4}{16\pi^2} 
\bigg\{f_0^4\left[\ln\left( \frac{2 \xi f_0^2\phi^2}{M^2}\right)-\frac{3}{2}\right] 
+5f_2^4\left[\ln\left( \frac{2 \xi f_2^2\phi^2}{M^2}\right)-\frac{1}{5}\right]\bigg\} \, .
\label{eq:V-symmetric}
\end{eqnarray}
The $\mO(\lambda^2)$ terms from scalar boson loops have been neglected.

In order to extract the minimum at $\tilde{\phi}=\langle\tilde{\phi}\rangle$ of the potential, we apply the critical point condition:
\begin{eqnarray}
&&\;\;\;\;\;\;\;\;\;\;\;\;\;\;
    \left. \frac{\dif V}{\dif \phi}\right|_{\langle\tilde{\phi}\rangle}\, =\, 0\, ,\qquad \left. \frac{\dif V}{\dif \theta}\right|_{\langle\tilde{\phi}\rangle} \, =\, 0 \,,
\end{eqnarray}
The first constraint yields exactly the same relation for $\lambda$ found in the one-scalar case in Eq.~(\ref{eq:1S-minimum-rel}).  
On the other hand, the second condition is trivially fulfilled, as the symmetric potential does not depend on $\theta$. 
Thus, we have a continuous set of non-trivial vacua with $\phi=\langle \phi\rangle\neq 0$ and any value of $\theta$, related through $U(1)$ transformations. This scenario is traditionally referred as spontaneous symmetry breaking and is characterised by a potential that is independent of the Goldstone fields:
\begin{equation}
    V  = \frac{\xi^2 \phi^4}{64\pi^2}\left[9\xi^2 f_0^4 + 4f_0^4 +20f_2^4\right]\left[\ln\left(\frac{\phi^2}{\langle\phi\rangle^2}\right)-\frac{1}{2}\right]\, .
\label{eq:V-symmetryc}
\end{equation}

This potential allows us to compute the masses of the scalar excitations around the chosen vacuum. In this case we have two degrees of freedom, so we have two scalar masses. The squares of these are the eigenvalues of the Hessian matrix at the minimum. In the $U(1)$ limit, the crossed elements of the matrix vanish and the physical eigenstates are $S=\phi-\langle\phi\rangle$ and the canonically normalized Nambu-Goldstone boson $\theta_c$, with masses,~\footnote{  
In order to compute the mass of the angular scalar component, we have to take into account that its kinetic term is given by $\frac{1}{2}\langle \phi\rangle^2 \partial_\mu\theta\partial^\mu\theta$. In order to have a canonically normalized kinetic term $\frac{1}{2} \partial_\mu\theta_c\partial^\mu\theta_c$, it is necessary to rescale the field in the following way: $\theta_c=\langle \phi\rangle \theta$.} 
\begin{eqnarray}
m_S^2&=&\left.\frac{\dif^2V}{\dif \phi^2}\right|_{\langle\tilde\phi\rangle }=\frac{\xi^2\langle\phi\rangle^2}{8\pi^2}\left(9\xi^2f_0^4+4f_0^4+20f_2^4\right) \,,
\nonumber\\
m_{\theta_c}^2&=&\left.\frac{1}{\langle \phi\rangle^2    }\frac{\dif^2V}{\dif \theta^2}\right|_{\langle\tilde\phi\rangle }=0\, .
\end{eqnarray}
Since we have assumed an exact $U(1)$ symmetry, $m_{\theta_c}^2=0$ and $\theta_c$ is an exact Nambu-Goldstone boson.

In the case of a small value of $\xi f_j^2$ and $\xi^2 f_j^2$ (with $j=0,2$), we could also consider another interesting scenario: if we gauge the $U(1)$ symmetry, the corresponding vector boson may trigger the spontaneous symmetry breaking in the fashion described in Sec.~\ref{sec:SQED-CW}.   
The effective potential gains then an additional contribution generated by the gauge boson loops ($e^4$ term in Eq.~\eqref{eq:Vgauge}). 
If, in addition to the basic CW assumption ($\lambda^2 \ll e^4$), the gravitational couplings are suppressed with respect to the $U(1)$ charge ($\xi^2 f_j^4,\, \xi^4 f_j^4\ll e^4 $, with $j=0,2$), then gauge boson loops dominate in the effective potential and scalar and gravitational loops can be neglected. 
We are led to the spontaneous symmetry breaking described in Sec.~\ref{sec:SQED-CW}, where the scalar field modulus $\phi$ gets a non-zero vev with value 
\begin{equation}\label{eq:vev_gauge}
    \langle\phi\rangle \, =\,  e^{-1} \, M \,  \exp\left[ \frac{1}{6} - \frac{4\pi^2}{9}\frac{\lambda}{e^4}\right] \ , 
\end{equation}
where scalar and gravitational loops have been neglected. Notice that this is the result obtained in Eq.~\eqref{eq:SQED-minimum}. 
This vev~\eqref{eq:vev_gauge} implies the appearance in the action of a Planck mass term given by $\bar{M}_{\text{Pl}}=\xi\langle\phi\rangle^2$, generated by the referred spontaneous symmetry breaking. Likewise, the Goldstone boson is no longer physical and it is absorbed by the $U(1)$ gauge boson, which becomes massive, with $m_A^2=e^2\langle\phi^2\rangle$.
Finally notice that, depending on the value of $\xi$ (i.e., $\xi\sim1$, $\xi\ll1$ or $\xi\gg1$), the pattern of the mass spectrum of the scalars, the massive spin--2 state, the massive $U(1)$ gauge boson and the Planck scale can vary. 
This mechanism will be treated in future works and we will not further discuss it here. In what follows we will not introduce this additional gauge boson, and we will just focus on the case where the spontaneous symmetry breaking is due to the gravitational interactions. 

\subsection{Soft $U(1)$ breaking} 

We will now consider that the $U(1)$ symmetry is not exact but approximate, with  $\xi_{1,2}\equiv \xi\pm\delta \equiv \xi(1\pm \hat{\delta})$. 
We will perform a perturbative expansion of the potential up to one loop in powers of $\hat{\delta}$ for $|\hat{\delta}|\ll 1$: 
\begin{eqnarray}
V(\phi,\theta)&=& 
V_0(\phi,\theta) +
\frac{9\xi^4 f_0^4\phi^4}{64\pi^2}(1+2\hat{\delta}\cos(2\theta))^2 
  \left[\ln\left((1+2\hat{\delta}\cos(2\theta))\frac{3\xi^2 f_0^2\phi^2}{M^2}\right)-\frac{3}{2}\right]
\nonumber\\
&&
   +\,\, \frac{\xi^2 \phi^4}{16\pi^2}(1+ \hat{\delta}\cos(2\theta))^2
\bigg\{   f_0^4\left[\ln\left((1+\hat{\delta}\cos(2\theta))\frac{2 \xi f_0^2\phi^2}{M^2}\right)-\frac{3}{2}\right] 
\nonumber\\
&&
+\,\, 5f_2^4\left[\ln\left((1+\hat{\delta}\cos(2\theta))\frac{2\xi f_2^2\phi^2}{M^2}\right)-\frac{1}{5}\right]\bigg\}
\, +\, \mathcal{O}(\hat{\delta}^2) \,, 
\end{eqnarray}
where we have neglected $\mO(\lambda^2)$ terms, and we can write the tree-level potential as 
\begin{eqnarray}
V_0(\phi,\theta) &=& \frac{\lambda}{4!}\phi^4 \, +\, \frac{\alpha}{4!}\phi^4 \cos(2\theta)  \, +\, \frac{\beta}{4!}\phi^4 \cos^2(2\theta)\,. 
\end{eqnarray}
We have made use of the following relations: 
\begin{eqnarray}
\xi_1^2\phi_1^2+\xi_2^2\phi_2^2&=&\xi^2(\phi_1^2+\phi_2^2)+2\delta\xi(\phi_1^2-\phi_2^2)+\mathcal{O}(\delta^2)
=\xi^2\phi^2 (1+2\hat{\delta}\cos(2\theta))+\mathcal{O}(\hat{\delta}^2) \, ,
\nonumber\\ 
\xi_1\phi_1^2+\xi_2\phi_2^2&=&\xi(\phi_1^2+\phi_2^2)+\delta(\phi_1^2-\phi_2^2)
=\xi\phi^2(1+\hat{\delta}\cos(2\theta)) \, .
\end{eqnarray}

We note that the terms of the tree-level potential absorb different one-loop gravitational UV divergences: $\lambda$ renormalizes the $\mO(\hat{\delta}^0)$ divergences; $\alpha$ cancels $\mO(\hat{\delta})$ divergences; and $\beta$ absorbs the $\mO(\hat{\delta}^2)$ UV infinities. 
According to the CW framework, the spontaneous symmetry breaking pops up due to an interplay between the tree-level potential of the scalar fields and the loop contributions with other particles, which are assumed to be of the same order around the vacuum of the theory. Hence, it sounds reasonable to assume the soft-breaking scalings $\alpha \sim \mO(\hat{\delta})$ and $\beta\sim \mO(\hat{\delta}^2)$ in our CW approach. 
The potential would be then arranged in the form 
\begin{equation}
V(\phi,\theta) = V_{(\hat{\delta}^0)}(\phi)\, +\, 
\hat{\delta}\, \cos(2\theta)\, g(\phi) \, +\, \mO(\hat{\delta}^2)\,,  
\end{equation}
with the symmetric potential $V_{(\hat{\delta}^0)}(\phi)$ provided by Eq.~(\ref{eq:V-symmetric}), and the first correction $\mO(\hat{\delta})$, given by
\begin{eqnarray} 
g(\phi) &=& 
\frac{\alpha}{4!\hat{\delta}} \phi^4
  +\frac{\phi^4}{16\pi^2}  
\bigg\{9\xi^4f_0^4  
  \left[\ln\left(\frac{3\xi^2f_0^2 \phi^2}{M^2}\right)-1\right] 
\\
&&
\hspace*{2cm}    +\,\,  2 \xi^2 f_0^4\left[\ln\left( \frac{2 \xi f_0^2\phi^2}{M^2}\right)-1\right]  
+10\xi^2 f_2^4\left[\ln\left( \frac{2 \xi f_2^2\phi^2}{M^2}\right)+\frac{3}{10}\right]\bigg\} \,.
\nonumber 
\end{eqnarray}
Here, we are assuming that $\beta\sim \mO(\hat{\delta}^2)$ as we have discussed above.
In the case when the symmetry breaking coupling $\beta$ scales like $\mO(\hat{\delta})$ one should consider the contribution of this term of the tree-level potential to $g(\phi)$. Our intention in this subsection is, however, to illustrate the soft symmetry breaking, so this $\beta$--term will be neglected.

We extract the minimum $\langle\tilde{\phi}\rangle$ of the potential $V(\phi,\theta)$ by demanding the critical point condition:
\begin{equation}
    \left. \frac{\dif V}{\dif \phi}\right|_{\langle\tilde{\phi}\rangle}\,=\, V_{(\hat{\delta}^0)}'(\langle\phi\rangle)\, +\, \mO(\hat{\delta})\, =\, 0\, ,\qquad \left. \frac{\dif V}{\dif \theta}\right|_{\langle\tilde{\phi}\rangle} \,=\, 
-\, 2 \hat{\delta}\, \sin(2\langle \theta\rangle)\, g(\langle\phi\rangle) \, +\, \mO(\hat{\delta}^2)\, =\, 0 \,,
\label{eq:2S-minimum-eq}
\end{equation}
where the first equation implies the same relation for $\lambda$ found in Eq.~(\ref{eq:1S-minimum-rel}) up to corrections $\mO(\hat{\delta})$. This substitution turns $V_{(\hat{\delta}^0)}(\phi)$ into the simplified result given in Eq.~(\ref{eq:V-symmetryc}), up to $\mO(\hat{\delta})$ corrections. 
The second equation in~(\ref{eq:2S-minimum-eq}) implies 
\begin{equation}
\sin\left(2\langle\theta\rangle\right)\, =\, 0 \quad \implies \quad \langle\theta\rangle\,=\,\frac{k \, \pi}{2} \, =\, 0,\, \frac{\pi}{2}, \, ... \qquad (k\in\mathbb{Z})
\end{equation}
Studying the Hessian, we find that if $\hat{\delta} g(\langle\phi\rangle)$ is positive (negative) the minimum is located at $\theta=\pi/2$ (at $\theta=0$), this is, at  $\langle\phi_1\rangle=0$ and $\langle\phi_2\rangle=\langle\phi\rangle$ (at $\langle\phi_1\rangle=\langle\phi\rangle$ and $\langle\phi_2\rangle=0$). We are just discussing the analysis of the first quadrant in the $(\phi_1,\phi_2)$ plane. The results are analogous for the other three quadrants. 

Finally, we compute the masses of the physical scalar excitations around the vacuum. There will be a massive radial component $S$, given by $\phi=\langle \phi\rangle +S +\mO(\hat{\delta})$, and a canonically normalized pseudo-Goldstone boson $\theta_c$, given by $\theta_c =\langle \phi\rangle \theta +\mO(\hat{\delta})$. Beyond the leading order in $\hat{\delta}$, we have a mixing between the $\phi$ and $\theta$ fields, which can be nevertheless ignored at lowest order. Thus, the mass of the physical eigenstates is given at the first non-vanishing order by
\begin{eqnarray}
m_S^2 &=& \left.\frac{\dif^2V}{\dif \phi^2}\right|_{\langle\tilde{\phi}\rangle}\, +\,\mO(\hat{\delta})\, =
V_{(\hat{\delta}^0)} ''(\langle\phi\rangle) \, +\, \mO(\hat{\delta}) \,, 
\nonumber\\
m_{\theta_c}^2&=&\left. \langle \phi\rangle^{-2} \frac{\dif^2V}{\dif \theta^2}\right|_{\langle\tilde{\phi}\rangle}  \, +\, \mO(\hat{\delta}^2)  
\, =\, 
\,   4 \langle \phi\rangle^{-2} \, \left|  \hat{\delta} \,\,  
 g(\langle\phi\rangle) \right| \, +\, \mO(\hat{\delta}^2) \, .
\end{eqnarray}
Employing the previous expressions for $V_{(\hat{\delta}^0)}(\phi)$ and $g(\phi)$, these expressions become:
\begin{eqnarray}
\label{finalmasses}
m_S^2 &=& \frac{\xi^2\langle\phi\rangle^2}{8\pi^2}\left(9\xi^2f_0^4+4f_0^4+20f_2^4\right) \, +\, \mO(\hat{\delta})
\,, 
\nonumber
\\
m_{\theta_c}^2&=&  |\hat{\delta}|\,\, \langle\phi\rangle^2 \, \,
\bigg| \frac{\hat{\alpha}}{6}  +  
\frac{9\xi^4 f_0^4 }{4\pi^2}\left[\log\left(\frac{3\xi^2 f_0^2\langle\phi\rangle^2}{M^2}\right) - 1\right]
\nonumber\\
&&\hspace*{0.5cm}
+\,\, \frac{\xi^2f_0^4 }{2\pi^2}\left[\log\left(\frac{2\xi f_0^2\langle\phi\rangle^2}{M^2}\right) - 1\right]
+ \frac{5\xi^2   f_2^4 }{2\pi^2}\left[\log\left(\frac{2\xi f_2^2\langle\phi\rangle^2}{M^2}\right) +\frac{3}{10} \right] \bigg|
 + \mathcal{O}(\hat{\delta}^2) \, ,
\nonumber
\\
\end{eqnarray}
with $\alpha\equiv \hat{\alpha}\hat{\delta}$, and the relation with reduced Planck mass given by $ 
\bar{M}_{\text{Pl}}=\xi\langle\phi\rangle^2  $~\cite{agravity,masaplanck}. 
Since we have slightly and explicitly broken the $U(1)$ global symmetry, the angular mode can be understood as a pseudo-Goldstone boson (even in the case with $\langle\theta\rangle=0$). This boson acquires a suppressed mass, $\mO(\hat{\delta})$, not only with respect to the Planck scale, but also to the mass of the radial scalar mode, following a hierarchy $m_{\theta_c}^2\ll m_S^2\ll \bar{M}^2_{\rm Pl}$.~\footnote{Read App.~\ref{app:Masses_SS}. In addition, notice that for $f_{0,2}\sim \mO(1)$ the masses $m_{0,2}^2= f_{0,2}^2 \bar{M}^2_{\rm Pl}/2$ of the extra massive spin-two and spin-zero mediators of the gravitational interaction become of the order of the Planck scale.}

\section{Conclusions}
\label{sec:conclusions}

In this work, we have carried out a study of the CW mechanism triggered by the gravitational interaction. First, we have reviewed the case studied by Coleman and Weinberg for scalar electrodynamics with quartic self-interaction.  
Then, we have extended the same mechanism for a renormalizable Lagrangian that couples a scalar field and gravity. It should be noted that despite being of different natures, there are great similarities between the electromagnetic and the gravitational interaction with respect to this mechanism since we have been able to extrapolate the procedure in a relatively simple way. The gravitational interaction is able to produce the same effect by providing a finite mass and a vev to the scalar field. 

The main difference, is that the CW mechanism produces the symmetry breaking of the gauge interaction for the electromagnetic case. Therefore, the photon acquires mass since a Higgs-like mechanism takes place. On the contrary, for the gravitational interaction, although the CW mechanism breaks scale invariance as in the electromagnetic case, it preserves general diffeomorphisms, which is the local symmetry associated to the interaction. Therefore, the standard spin-two graviton remains massless. In any case, the dimensional transmutation of the CW mechanism induces the Planck scale in a natural way. 

Finally, we have extended our study for the gravitational CW mechanism to a multi-scalar case. In particular, we have analysed in detail a model with two scalar degrees of freedom. We pay special attention to the case of an approximated global $U(1)$ symmetry in the scalar sector. This has allowed us to study the system by performing a perturbative expansion in terms of a $\hat{\delta}$ parameter characterising the explicit $U(1)$ symmetry breaking.  
If we consider equal coupling constants for the two scalar fields, the global $U(1)$ symmetry is exact and the theory contains a massive scalar and an exact Goldstone boson with zero mass. Deviations from this limit turns this state into a pseudo-Goldstone with a non-vanishing square mass proportional to the soft breaking parameter $\hat{\delta}$. The study of a general multi-scalar case can be performed in an straightforward way by generalising the studies performed in this analysis, leading to similar qualitative conclusions. 
This type of models provide a viable construction of a pseudo-Goldstone sector with masses at a scale much lower than the Planck scale. These ideas can be pursued in order to introduce the Higgs action and the Standard Model within this framework.

\acknowledgments

This work was supported by Grants FPU16/06960 (MECD), the MICINN (Spain) project PID2019-107394GB-I00/AEI/10.13039/501100011033 (AEI/FEDER, UE), and PID2019-108655GB-I00/AEI/10.13039/501100011033, UCM under research group 910309, IPARCOS institute, and by EU STRONG-2020 under the program H2020-INFRAIA-2018-1 [grant no. 824093]. JARC acknowledges support by Institut Pascal at Université Paris-Saclay during the Paris-Saclay Astroparticle Symposium 2022, with the support of the P2IO Laboratory of Excellence (program “Investissements d’avenir” ANR-11-IDEX-0003-01 Paris-Saclay and ANR-10-LABX-0038), the P2I axis of the Graduate School of Physics of Université Paris-Saclay, as well as IJCLab, CEA, APPEC, IAS, OSUPS, and the IN2P3 master projet UCMN. Part of the contribution of JARC to this article is based upon work from COST Action COSMIC WISPers CA21106, supported by COST (European Cooperation in Science and Technology).

\appendix

\section{Projectors, tensor relations and traces}
\label{app:projectors}

For the longitudinal and transverse projectors we have the relations 
\begin{eqnarray}
T\cdot T=T\,, \qquad  L\cdot L =L\, , \qquad L\cdot T=T\cdot L =0\, , 
\nonumber\\
 T_{\mu\nu}+L_{\mu\nu}=\eta_{\mu\nu}\, ,\quad {\rm Tr}T=(D-1)\, ,\qquad {\rm Tr}L=1\,   .
\end{eqnarray}
Notice that, for $D=4$, the last relations become ${\rm Tr}\,T=3$ and ${\rm Tr}\,L=1$

For the graviton propagator projectors, we have analogous relations:
\begin{eqnarray}
&& P^{(i)}\cdot P^{(i)} =P^{(i)}\, ,\quad 
P^{(i)}\cdot P^{(j\neq i)} = P^{(j\neq i)}\cdot P^{(i)} = 0\, ,
\nonumber 
\\
&&\hspace{1cm} \sum_{i=0w,0,1,2} P^{(i)}_{\mu\nu,\alpha\beta} 
=\frac{1}{2}\left(\eta_{\mu\alpha}\eta_{\nu\beta}+\eta_{\mu\beta}\eta_{\nu\alpha}\right) \,,
\nonumber
\\
&&\hspace{-1cm} {\rm Tr}P^{(2)}=\frac{(D^2-D-2)}{2}\, ,\quad {\rm Tr}P^{(1)}=D-1\, , \quad {\rm Tr}P^{(0)}=1\, ,\quad {\rm Tr}P^{(0w)}=1\, , 
\end{eqnarray}
where here, there is not implicit sum over repeated indices unless explicitly stated. 
Notice that, for $D=4$, the last relations become ${\rm Tr}P^{(2)}=5$, ${\rm Tr}P^{(1)}=3$,  ${\rm Tr}P^{(0)}=1$ and ${\rm Tr}P^{(0w)}=1$.

Here we provide some intermediate steps of the calculation of the gravity loops:
\begin{eqnarray}
D\cdot C_{(2\xi)} &=& \frac{2\xi}{k^2} \left(\left(\frac{(D-2)}{2} f_0^2\right) P^{(0)} \, +\, f_2^2 P^{(2)}\right)\, ,   
\\
\left(D\cdot C_{(2\xi)}\right)^n &=& \left(\frac{2\xi}{k^2}\right)^n  \left(\left(\frac{(D-2)}{2} f_0^2\right)^n   P^{(0)} \, +\, f_2^{2n} P^{(2)}\right)\, ,    
\nonumber\\
{\rm Tr}\left\{ \left(D\cdot C_{(2\xi)}\right)^n\right\} &=& \left(\frac{2\xi}{k^2}\right)^n  \left(\left(\frac{(D-2)}{2} f_0^2\right)^n   \times  \underbrace{ 1}_{={\rm Tr}P^{(0)}} \, +\, f_2^{2n} \times \underbrace{\left(\frac{D^2-D-2}{2}\right)}_{={\rm Tr} P^{(2)}}\right)\, .  
\nonumber 
\end{eqnarray}

\section{Masses of the scalar sector}
\label{app:Masses_SS}

In this type of models, the scalar sector mixes due to the presence of finite vevs in the non-minimal gravitational couplings. For instance, in the case of a single field, the normalized quadratic term in the action can be written in terms of the matrix:
\begin{eqnarray}
\label{Hessian2}
 &&
    M_{2}=
    \begin{pmatrix}
    \frac{3}{32} \left(\frac{\partial_\mu \partial^\mu}{m_0^2}+1\right)\partial_\nu \partial^\nu
    & \frac{3}{4} \xi^{\frac{1}{2}}   
    \partial_\mu \partial^\mu \\
    \frac{3}{4} \xi^{\frac{1}{2}}    
     \partial_\mu \partial^\mu 
    & - \left(\partial_\mu \partial^\mu +m_S^2\right)
    \end{pmatrix}\,,
\end{eqnarray}
associated to the scalar degree of freedom of the metric perturbation and the scalar mode corresponding to the field that develops the vev, as it has been discussed in \cite{Kubo:2022dlx}. In the limit
where $\xi\longrightarrow 0$, the two scalar modes do not mix, and the square mass eigenvalues are just 
$m_0^2$ and $m_S^2$. However, $\xi$ does not need to be small. In such a case, the square mass eigenvalues corresponding to the propagating scalar eigenstates can deviate strongly from the above values:
\begin{eqnarray}
    m^2_{\text{heavy}}&=&(1+6\xi)\,m_0^2 \left(1+\frac{6\xi}{(1+6\xi)^2}\frac{m_S^2}{m_0^2}+\mO\left(\frac{m_S^4}{m_0^4}\right)\right)\,,\nonumber \\
    m^2_{\text{light}}&=&\frac{1}{(1+6\xi)}\,m_S^2\left(1-\frac{6\xi}{(1+6\xi)^2}\frac{m_S^2}{m_0^2}+\mO\left(\frac{m_S^4}{m_0^4}\right)\right) \,.
\end{eqnarray}
In any case, the perturbativity and the CW condition of the model forces to have a well-defined hierarchy between the two masses: $m^2_{\text{heavy}}>m_0^2\gg m_S^2>m^2_{\text{light}}$. 
Note that the recovery of the Planck scale by  $\bar{M}_{\rm pl}^2=\xi\langle\varphi\rangle^2$ requires $\xi>0$.
This result is consistent with previous studies within the same Jordan frame~\cite{Kubo:2022dlx} or the Einstein frame~\cite{agravity}.

For a gravitational CW mechanism with several scalar fields, the situation is more involved. In general, there will be mixing terms due to the non-minimal coupling, between the scalar mode of the metric and each additional scalar field that develops a vev. Therefore, the Planck mass will have contributions from each of these vevs. However, an important simplification arises if there is only one scalar field with non-minimal coupling or only one scalar field acquires a vev. Under these conditions, only this field mixes with the scalar mode of the geometry. In such a case, the propagation matrix~\eqref{Hessian2} is trivially extended without additional non-diagonal entries. This is exactly what happens for the case with two additional scalar degrees of freedom with an exact global $U(1)$ symmetry discussed in this work. The radial mode mixes with the scalar graviton, whereas the angular mode remains as a propagation eigenstate with zero mass. Indeed, when the $U(1)$ symmetry is slightly broken, the situation does not change at first order in the soft breaking parameter $\hat{\delta}$, and the mass $m_{\theta_c}^2$ given by Eq. \eqref{finalmasses} is not modified at leading order.

\bibliographystyle{apsrev4-1}
\bibliography{bibliografia}

\end{document}